\documentclass{frontiersSCNS} 

\usepackage{url,lineno}

\usepackage{setspace}
\usepackage[figuresright]{rotating} 

\usepackage{graphicx}
\usepackage{placeins}

\usepackage{rotating}
\usepackage{tikz}
\usepackage{pdflscape}
\usepackage{eqnarray}

\copyrightyear{}
\pubyear{}

\def\journal{Psychology}
\def\DOI{}
\def\articleType{Research Article}
\def\keyFont{\fontsize{8}{11}\helveticabold }

\def\firstAuthorLast{C. Alonso-Montes {et~al.}} 
\def\Authors{Carmen Alonso-Montes\,$^{1,*}$, 
Ibai Diez\,$^{2}$, 
Lakhdar Remaki\,$^{1}$, 
I\~{n}aki Escudero\,$^{2,3}$, 
Beatriz Mateos\,$^{2,3}$, 
Yves Rosseel\,$^{4}$, 
Daniele Marinazzo\,$^{4}$, 
Sebastiano Stramaglia\,$^{1,5}$, 
and Jesus M. Cortes,$^{2,6,7}$}
\def\Address{$^{1}$BCAM - Basque Center for Applied Mathematics, Bilbao, Spain \\
$^{2}$ Biocruces Health Research Institute, Cruces University Hospital.	 Barakaldo,	Spain \\
$^{3}$Radiology Service,	Cruces University Hospital. Barakaldo, Spain\\
$^{4}$Department of Data Analysis, Faculty of Psychological and Pedagogical Sciences, Ghent University, Belgium\\
$^{5}$Dipartimento di Fisica, Universit\'a degli Studi di Bari and INFN. Bari, Italy\\
$^{6}$Ikerbasque, The Basque Foundation for Science. Bilbao, Spain.	  \\
$^{7}$Department of Cell Biology and Histology. University of the Basque Country. Leioa, Spain
}
\def\corrAuthor{Carmen Alonso-Montes}
\def\corrAddress{BCAM - Basque Center for Applied Mathematics, Alameda Mazarredo 14, E-48009, Bilbao, Spain}
\def\corrEmail{calonso@bcamath.org}

\begin{document}
\onecolumn
\firstpage{1}

\title[eSEM vs SC  in the resting brain]{Lagged and instantaneous dynamical influences related to brain structural connectivity}
\author[\firstAuthorLast ]{\Authors}
\address{}
\correspondance{}
\extraAuth{}
\topic{}

\maketitle


\clearpage

\begin{abstract}
Contemporary neuroimaging methods can shed light on the basis of human neural and cognitive specializations, with important implications for neuroscience and medicine. 
 Indeed, different MRI acquisitions  provide different brain networks at the macroscale; whilst diffusion-weighted MRI (dMRI) provides a structural connectivity (SC) coincident with the bundles of parallel fibers between brain areas, functional MRI (fMRI) accounts for the variations in the blood-oxygenation-level-dependent T2* signal, providing functional connectivity (FC).  
Understanding the precise relation between FC and SC, that is, between brain dynamics and structure, is still a challenge for neuroscience.
To investigate this problem,  we acquired data at rest    and built the corresponding SC (with
 matrix elements corresponding  to the fiber number between brain areas) to be compared  with FC connectivity matrices obtained by three different methods: directed dependencies by an exploratory   version of  structural equation modeling (eSEM), linear correlations (C) and partial correlations (PC). We also considered the possibility of using lagged correlations in time series; in particular, we   compared    a lagged version of eSEM   and  Granger causality (GC).  Our results were two-fold: firstly, eSEM performance in correlating with  SC was comparable to those obtained from C and PC, but eSEM (not C, nor PC) provides information about directionality of the functional interactions. Second, interactions on a time scale much smaller than the sampling time,  captured by instantaneous connectivity methods, are much more related to SC than slow directed influences captured by the lagged analysis. Indeed the performance in correlating with  SC was much worse for GC and for the lagged version of eSEM. We expect these results  to supply further insights to the  interplay between SC and functional patterns, an important issue in the study of   brain physiology and function.

\tiny
 \keyFont{ \section*{Keywords:} structural equation modeling; functional connectivity; structural connectivity; resting state; functional magnetic resonance imaging; tensor diffusion imaging  
 }
\end{abstract}


\clearpage
 \section*{List of Acronyms}
\begin{itemize}
\item[] SC: Structural Connectivity
\item[] FC: Functional Connectivity
\item[] EC: Effective Connectivity
\item[] eSEM: Exploratory Structural Equation Modeling (a non-lagged model)
\item[] eSEM1: eSEM with lag=1
\item[] eSEM2: eSEM with lag=2
\item[] eSEM3: eSEM with lag=3
\item[] GC: Granger Causality
\item[] GC1: GC with lag=1
\item[] GC2: GC with lag=2
\item[] GC3: GC with lag=3
\item[] C: Correlation
\item[] PC: Partial Correlation
\item[] RSN: Resting State Network
\item[] DMN: Default Mode Network
\item[] SM: Sensory Motor
\item[] ExC: Executive Control
\item[] ROI: Region of Interest
\item[] CP: (Structurally) Connected Pairs
\item[] NCP: (Structurally) Non-Connected Pairs
\end{itemize}

\clearpage

\section{Introduction}

Three different main  classes of brain networks  are currently investigated  \citep{Sporns2005,Bonifazi2009,Sporns2004,Friston1994,Friston2011}: networks defined by their “structural connectivity” (SC) refer to anatomical connections between brain regions; networks defined by their “functional connectivity” (FC) account  for statistical similarities in the dynamics  between   distinct neuronal populations; and “effective connectivity” (EC) networks identify  interactions or  information flow between regions.

Current   magnetic resonance imaging (MRI) techniques have allowed SC, FC and EC  brain networks to be measured at the macroscale.  Thus, SC networks have been obtained from diffusion tensor images (DTI) and high-resolution tractography \citep{Craddock2013} while   FC networks have been obtained   from  correlations between blood oxygen-level dependent (BOLD) time-series \citep{Biswal1995}.  

Different methods can assess EC.     One possibility  is the dynamic causal modeling, addressing how the activity in one brain area is affected by the activity in another area using explicit models of neural populations \citep{Friston2003,Penny2004}. Other possibilities are  data-driven approaches with no further assumptions  about the hemodynamic response, nor about the biophysics  of the BOLD signal from individual neuron to  population level.  Two popular existing data-driven methods  
to calculate EC are  Granger causality (GC) \citep{granger1969} and transfer entropy \citep{Schreiber2000}. 

Another well-known method to  calculate EC is the    structural equation modeling (SEM). Although SEM  assumes an  implicit  model (ie. an influence matrix) \citep{Bollen1989},  in the present study we focus on  an exploratory version of SEM   (labelled eSEM) where  all variables might  (a priori)   interact  with all the others. 
Notice that,  eSEM   is by construction   exploratory whilst  SEM is largely confirmatory. Both SEM and eSEM are methods to calculate EC,  
since both methods provide directed connectivity matrices. 

In this paper, we aim   to bring some  light    in a long lasting   question: How brain structure is shaped by its function, and viceversa? Or alternatively, using the   language of networks:  How are the three classes of networks SC, FC and EC   related to each other?  It is important to emphasize that, this  challenging problem has not yet a clear answer    for any general brain condition and data set. Here,  to address this question,
we will focus here  in     the resting brain, i.e. when the brain   is not performing any goal-oriented task.

Notice that despite the simplicity of the context where 
these patterns of brain activity are generated,  the resting brain dynamics is complex, encompassing a superposition of multiple  resting state networks (RSNs) ~\citep{Raichle2001,Fox2005,Raichle2006,Raichle2007,Raichle2009};   each  RSN underlying a different cognitive function  e.g., there are visual networks, sensory-motor networks, auditory networks, default mode networks, executive control networks, and some others (for further details see for instance ~\citep{Beckmann2005} and references therein).

 Pioneering work showed        that  SC and FC are correlated to some extent   \citep{Hagmann2008,Honey2009}.  After these fundamental papers,  some other    studies     made use of the  combined data sets to address different aspects of brain dynamics  ~\citep{Fraiman2009,Cabral2011,Deco2011,Haimovici2013,Marinazzo2014,Messe2014,Goni2013,Kolchinsky2014}.  
In this paper, both structural and functional data have been used to demonstrate to which extent   the EC obtained by eSEM and the FC obtained by C and PC  are similar to SC.  

Previous approaches analyzed  fMRI data based on SEM ~\citep{bullmore2000,Schlosser2006,kim2007,gates2010,gates2011}, dealing  with subsets of candidate regions selected on the basis of prior knowledge. However, the performance of these approaches depends strongly on the correctness and completeness of the hypothesized model of connections. In the present work, eSEM is applied in an exploratory fashion  to a multivariate dataset corresponding to a specific brain system consisting in 15 different regions of interest (ROIs), fully covering (with no  further 
assumptions about the underlying connectivity) three of the well-known RSNs ~\citep{Beckmann2005}: The sensory-motor network (SM), the executive-control network (ExC) and the default mode network (DMN). The application of   eSEM   returns an influence matrix which is not symmetric (ie., a region A can influence  B differently than how B influences A) and describes  fully connected directed dependencies between ROIs.

The performance  achieved by eSEM  in correlating with SC (thus, measusing the similarity between eSEM and SC) is also  compared with FC, obtained by two other methods: the linear correlation (C) and partial correlation (PC). Unlike C, PC is commonly used to analyze  direct relationships among fMRI time series with good performance~\citep{Marrelec2006,Marrelec2009,Maki-Marttunen2013}, since network influences beyond the specific pair are removed. 

Furthermore,  eSEM is also applied to lagged time series to estimate a saturated, fully connected, but recursive model.  Notice that bi-directional influences here are detected as cross-lagged effects.  The results from this {\it lagged} version of eSEM are compared with those from   GC.  As a result, we will show that lagged methods   are  less related to SC measures, which  implies that 
the dependencies found in the data on slower time scales (in comparison to instantaneous interactions) are less related to SC.

\section{Material \& Methods}

\subsection{Same-subject structure-function acquisitions} 
This work was approved by the Ethics Committee at the Cruces University Hospital; all the methods were carried out in accordance to approved guidelines. A population of n=12 (6 males) healthy subjects, aged between 24 and 46 (33.5 $\pm$ 8.7), provided information consents   before the   imaging session. For all the participants, we acquired same-subject structure-function data with a Philips Achieva 1.5T Nova scanner. The total scan time for each session was less than 30 minutes and high-resolution anatomical MRI was acquired using a T1-weighted 3D sequence with the following parameters: TR = 7.482 ms, TE = 3.425 ms; parallel imaging (SENSE) acceleration factor=1.5; acquisition matrix size=256x256; FOV=26 cm; slice thickness=1.1 mm; 170 contiguous sections. Diffusion weighted images (DWIs) were acquired using pulsed gradient-spin-echo echo-planar-imaging (PGSE-EPI) under the following parameters:  32 gradient directions, TR = 11070.28 ms, TE = 107.04 ms, 60 slices with thickness of 2 mm, no gap between slices, 128x128 matrix with an FOV of 23x23 cm. Changes in blood-oxygenation-level-dependent (BOLD) T2* signals were measured using an interleaved gradient-echo EPI sequence. The subjects lay quietly for 7.28 minutes, during which 200 whole brain volumes were obtained under the following parameters: TR = 2200 ms, TE = 35 ms; Flip Angle 90, 24 cm field of view, 128x128 pixel matrix, and 3.12 x 3.19 x 4.00 mm voxel dimensions.

We have shown in \citep{diez2015} that the relationship between SC and FC found  with the data used in this study is confirmed by the MGH-USC Human Connectome Project, of much higher quality. The results we show here    open the possibility to a generalization to many other data sets.

\subsection{Data preprocessing} 

\subsubsection{Structural data:} To analyze the diffusion  images (dMRI), the eddy current correction was applied to overcome artifacts produced by changes in the gradient field directions of the MR scanner and subject head movement.    In particular,  the  eddy-correct tool from FSL was used to correct both eddy current distortions, and simple head motion, using affine registration to a reference volume. After this,  DTIFIT was used to perform the fitting of the diffusion tensor for each voxel, using as an input the eddy-correct output.
No extra de-noising was applied in the data  and our results were not wrapped to any template.
Two computations were performed to transform the atlas to each individual space: (1) the transformation between the MNI template to the subject structural image (T1), and (2) the transformation between the T1 to the diffusion image space.
Combining both transformations, each atlas region is transformed to the diffusion space, allowing
to count the number of fibers connecting all ROIs pairs.    Using the corrected data, a local fitting of the diffusion tensor was applied to compute the diffusion tensor model at each voxel. Then, a  deterministic tractography algorithm (FACT) \citep{Mori1999} was applied using TrackVis \citep{Wang2007}, an interactive software for fiber tracking.  

\subsubsection{Functional data:} The functional MRI (fMRI) data was preprocessed with FSL (FMRIB Software Library v5.0). The first 10 volumes were discarded for correction of the magnetic saturation effect and for the remaining volumes, first the movement is corrected 
 and then, the slice-time is also corrected for temporal alignment. All voxels were spatially smoothed with a 6mm FWHM isotropic Gaussian kernel and after intensity normalization, a band pass filter was applied between 0.01 and 0.08 Hz~\citep{Cordes2001}. Finally, linear and quadratic trends were removed. We next regressed out the motion time courses, the average CSF signal, the average white matter signal and the average global signal. Finally,   fMRI data was transformed to the MNI152 template, such that a given voxel had a volume of 3mm*3mm*3mm.    

It is important to emphasize that to remove or not the average global signal in
 FC studies is currently a controversial issue \citep{saad2012}; see also \url{http://rfmri.org/GSRDiscussion}. Here, following most of the studies addressing brain FC, we have applied   the global signal removal; but the situation of not applying the  global signal removal has been also explored  (figure \ref{figS2}).

\subsubsection{HRF blind deconvolution:}
In order to eliminate the confounding effect of HRF on temporal precedence, we individuated point processes corresponding to signal fluctuations with a given signature and extracted a voxel-specific HRF to be used for deconvolution, after following an alignment procedure. The parameters for blind deconvolution were chosen with a physiological meaning according to ~\citep{Wu2013}: for a TR equal to 2.2s, the threshold was fixed to 1 SD (standard deviation) and the maximum time lag was fixed to 5 TR (for further details on the complete HRF blind deconvolution method and the different parameters to be used, see ~\citep{Wu2013}). The resulting time-series, after   HRF blind deconvolution, are the ones used for the calculation of EC and FC.

\subsection{ROIs extraction}

Regions of interest (ROIs) were defined  by  using the masks of the resting state networks (RSNs)   reported in \citep{Beckmann2005}, which
can be downloaded from \url{http://www.fmrib.ox.ac.uk/analysis/royalsoc8/}. Note that, we are not dealing with the independent components per se, but with    the voxels time-series localized within the masks.    Similar approaches based on the RSNs masks to define ROIs  have been widely used before  \citep{Tagliazucchi2012, Haimovici2013, Carhart2014,Tagliazucchi2014, IF2015}.   

\begin{figure}[htb]
\center{
\includegraphics[width=14cm]{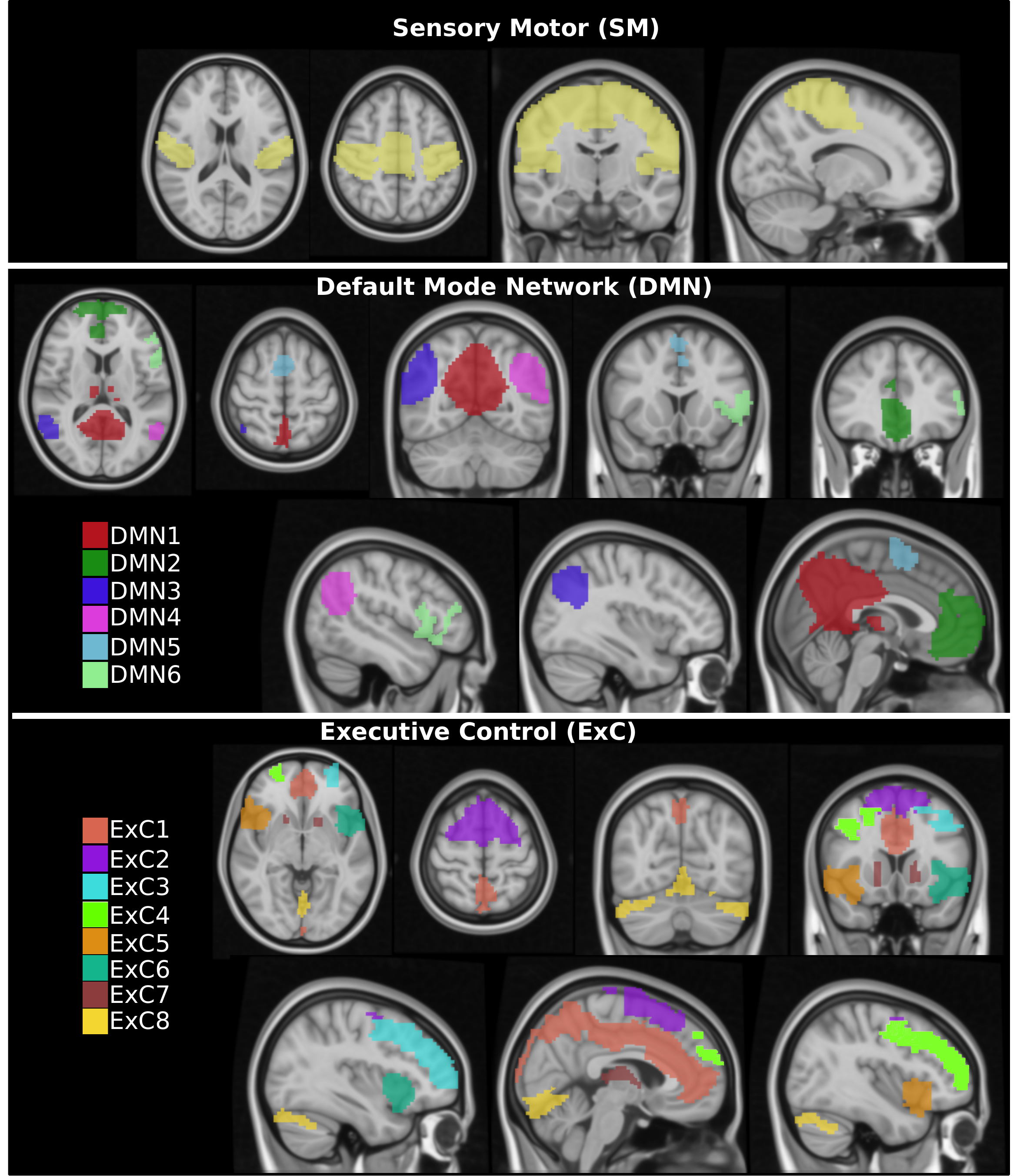}
\caption{\label{fig1} {\bf Sketch for regions of interest (ROIs).} 15 different ROIs were extracted from three different resting state networks:  1 ROI in the     sensory motor (SM), 6 ROIs in the default mode network (DMN) and 8 ROIs in the  executive control (ExC). }}
\end{figure}

Specifically, the following three RSNs were selected: the default mode network (DMN), the executive control (ExC) network and the sensory motor (SM) network.  Next, these three networks were manually    subdivided in distinct spatially contiguous regions (see figure 1). For each region, a “region growing” segmentation method was applied by manually selecting  a seed region,  thus obtaining a total of 15 different ROIs: 1 SM region, 6 DMNs and 8 ExCs regions. In particular, the "island effect" method   incorporated in 3D Slicer (\url{http://www.slicer.org}) was applied, which selects all the voxels of the contiguous region given an initial seed.  Visual representations of all ROIs are given in  figure \ref{fig1} and their sizes in      table \ref{tab1}.


\subsection{Calculation of structural,  functional and effective connectivity matrices}

\subsubsection{ Structural connectivity (SC):} Matrices were
obtained per each subject by counting the number of fibers connecting two ROIs (that is, starting in  one ROI and finalizing in another) for each individual pair;  thus, for a number of 15 ROIs, it gave 105  different values. 

\subsubsection{ Functional connectivity (FC): } Matrices were calculated by applying   to   the rs-fMRI time
series two methods: the  linear correlation coefficient (C) and the partial correlation analysis (PC). Here, C was  calculated by using the \textit{corr} function from Matlab (MathWorks Inc., Natick, MA).   Assuming C to be a non-singular  matrix,  the elements of the PC matrix satisfy  that $\mathrm{PC}_{ij} \propto   (C^{-1})_{ij} $,  so they are proportional to     the elements of the so-called precision matrix   \citep{Maki-Marttunen2013}. Here, PC was computed using the \textit{partialcorr} function from Matlab (MathWorks Inc., Natick, MA).
Thus,  PC  is an extension of C to calculate direct interactions between pairs,  as it achieves to remove  for a given  pair the correlation contribution from other pairs.

\subsubsection{Effective connectivity (EC)  by  the exploratory  structural equation modeling (eSEM).}  This  refers to a statistical technique aiming to estimate  Granger-causal  relationships based on quantitative and qualitative causal information, by means of linear regression-based models. Unlike regression, SEM is  formulated as a confirmatory model rather than a   predictive model. Being interested in the description of the directed dependencies between the 15  ROIs, avoiding any prior
hypothesis on the connectivity pattern, we here applied multiple regressions among all the variables. Therefore, our analysis by SEM has neither structural model nor a measurement model, and provides a fully connected estimate of the directed dependencies among all the pairs of ROIs. This exploratory analysis is referred as eSEM. This model   does not use temporal correlations in the data and it is applied to non-lagged time series. 

To estimate the model parameters of eSEM, a standard maximum likelihood  
estimation was used using the lavaan package   in R \citep{R2014, Rosseel2012}.   Notice that, this is justified since for saturated linear models, the maximum likelihood estimates   are identical to least squares  estimates.

In a second part of this study,  eSEM was also applied to lagged time series   to estimate a saturated, fully-connected, but recursive model. Notice that lagged eSEM is recursive but the non-lagged eSEM is not. The observed variables are the time series for the 15 ROIs augmented with lagged versions of the same time series.  For eSEM1, only the   time series accounting for lag=1 were added, resulting in 30 variables in total; for  eSEM2, both  lag =1 and lag =2 time series were added, resulting in 45 observed variables in total; finally, for eSEM3, lag=1, lag=2 and lag=3 time series were added, resulting in 60 observed variables in total. Three types of parameters were included in the model: (1) all autoregressive regressions  within each ROI   to take into account the time-dependencies; (2) all possible cross-lagged regressions between the ROIs; (3) (residual) covariances for all other pairwise relations that were not included in the set of regressions (for example, all contemporaneous connections).  Importantly,  contemporaneous regressions between ROIs at the same time point were not included. Moreover, to estimate the model parameters of eSEM1, eSEM2 and eSEM3, standard maximum likelihood estimation was used using the lavaan package \citep{Rosseel2012}.

After estimation of all model parameters, an influence matrix was computed as follows: For each pair, the evidence for this particular (directed) connection was collected. For eSEM1, this was simply the regression coefficient corresponding to the cross-lagged effect of one ROI on another (controlling for both auto-regressive effects and cross-lagged effects of other ROIs). That is, the effect of a ROI on the previous time point on a target ROI at the current time point. For eSEM2, this was a function (here, the product) of two regression coefficients: one for the effect of a ROI on the previous time point on the target ROI at the current time point (just like eSEM1), and one for the effect of a ROI measured two time points towards the target ROI at the current time point. This was done for all possible pairs, averaging  all ROIs of the influence matrix except for the diagonal, which was kept at zero.

Notice that the cross-lagged evidence is only used to determine the directed influence of one ROI on another, while controlling for both auto-regressive effects and the cross-lagged effects of other ROIs. In fact, the regression coefficients computed by eSEM1, eSEM2 and eSEM3 are identical to those that would be computed when  Granger causality (GC1, GC2, GC3, of order 1,   2  and 3 respectively) is employed \citep{granger1969}; see also suppl. material for further details.   But instead of computing an $F$-statistics for each pairwise connection as GC does, here, we use the product regression coefficient(s)   to average the influence matrix.

\subsection{Statistical Analysis}

The  values of  the average matrices across subjects  eSEM, C and PC  were compared into two  groups: values associated to structurally connected pairs (CP), meaning that  two ROIs are connected with a non-zero fiber number, and those ones associated to non-connected pairs (NCP), i.e., zero fibers existed between the two ROIs. A one-way ANOVA test was performed using the MATLAB function anova1 (MathWorks Inc., Natick, MA) between CP and NCP  (statistical significance is considered to have a p-value $<$ 0.01).  Thus, small p-values show that the connectivity matrices calculated on the two groups CP and NCP   have a different mean, ie., they are different from each another which indicates  that a given method can separate connected pairs from non-connected ones. The same analysis was also applied to  eSEM1, eSEM2, eSEM3, GC1, GC2, GC3.

\section{Results}

Firstly, the three different resting networks SM, DMN and ExC were selected.  Next,  the three networks were divided in a  total number of 15 different ROIs (see methods and figure \ref{fig1} for details).

\begin{figure}[htb]
\center{\includegraphics[width=14cm]{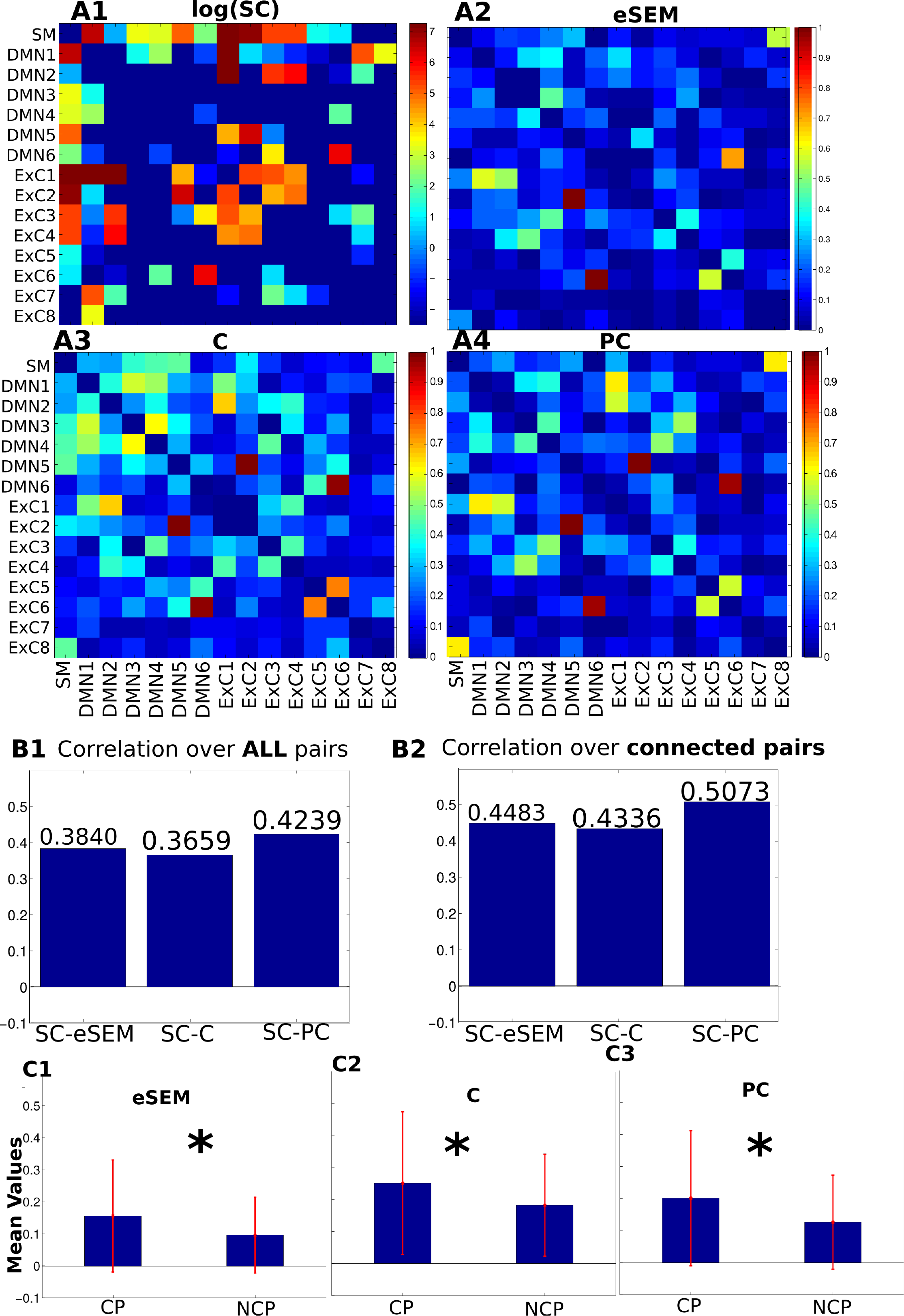}
\caption{\label{fig2} {\bf Structural, effective and functional connectivity matrices (SC,EC and FC, respectively)}:
{\bf A$_{1}$:} SC matrix calculated  by the fiber number. Because many of the values in this matrix are very small, we plotted it in logarithmic scale only to enhance visibility. {\bf A$_{2}$-A$_{4}$:} EC (eSEM) and FC   matrices (C and PC), all of them normalized in the  $[0,1]$ range  for comparison purposes.    
{\bf B:} Correlation-based similarity between SC and eSEM, C and PC, calculated either over all pairs or only on connected pairs.
{\bf C:} Mean values of connectivity matrices separated in two groups: pairs such that they have non-zero fibers between them (structurally connected pairs,  CP) and non-connected pairs (NCP). * $p-value< 0.01$, otherwise means no statistical significance.}}
\end{figure}

Next, the average across   subjects  SC matrix  (figure \ref{fig2}A$_1$) was computed by averaging the  fiber number between pairs of ROIs.  Notice that  SC  is a matrix with many near-zero values. So, it is represented in logarithmic scale just to improve visualization, but all the analyses were performed using the original SC matrix.

Next,  three   connectivity matrices were calculated for each subject from the rs-fMRI time series: eSEM,  C and PC (details in Methods).  Next, an  average  matrix across subjects  was calculated for all matrices.   The values of eSEM, C and PC after  normalization in the range [0,1] are represented in figures  \ref{fig2}A$_2$-A$_4$. Notice that, unlike C and PC, eSEM provides a non-symmetrical connectivity matrix.

To address the similarity   between these matrices,   and following previous work \citep{Hagmann2008,Honey2009},  the Pearson's correlation between  the SC   entries (vector-wise  using  all matrix elements) and the corresponding ones   for eSEM, C and PC was computed. 
The three connectivity matrices   increased their similarity (based on correlation) with SC  on connected pairs, pairs connected with non-zero fibers between ROIs, compared to the situation when all  pairs were used for the correlation calculation, ie., values in figure \ref{fig2}B$_2$ are bigger than in figure \ref{fig2}B$_1$.   The same results   also hold  when    Spearman's correlations were calculated (figure \ref{figS1}). It is important to emphasize that these results did not depend on the effect of removing or not the global signal to the time series data. Indeed, similar results than in  figures  \ref{fig2}B$_1$ and  \ref{fig2}B$_2$ were obtained without global signal removal (figure \ref{figS2}). Thus, after this  simple analysis, we show that   the three measures (eSEM, C and PC) are dependent on SC.

We next investigated whether average values of eSEM, C and PC had  significant differences between  CP and NCP  (non-connected pairs).   The three connectivity matrices showed   bigger     (significant)    values on CP compared to NCP (figure \ref{fig2}C), thus indicating that the three methods eSEM, C and PC separated the groups of structurally connected  links from those which were not connected. Moreover, PC performed better than eSEM, whilst eSEM and C performed approximately equal. 

Next, we addressed the effect that lagged interactions had on eSEM. Thus, when calculating eSEM on lagged-time series, eSEM could not distinguish (ie., the p-value between the two groups was high) between  CP and NCP (see figure \ref{fig5}). And this occurred independently on using eSEM or a different model accounting for lagged interactions, here,  the method of   multivariate GC was used (figure \ref{fig5}).
   These results  indicated that instantaneous measures of interactions (ie., approaches dealing only with equal-time correlations) are better shaped by SC in comparison to algorithms using temporal information (and this was observed both using eSEM and GC).

\begin{figure}[htb]
\center{\includegraphics[width=14cm]{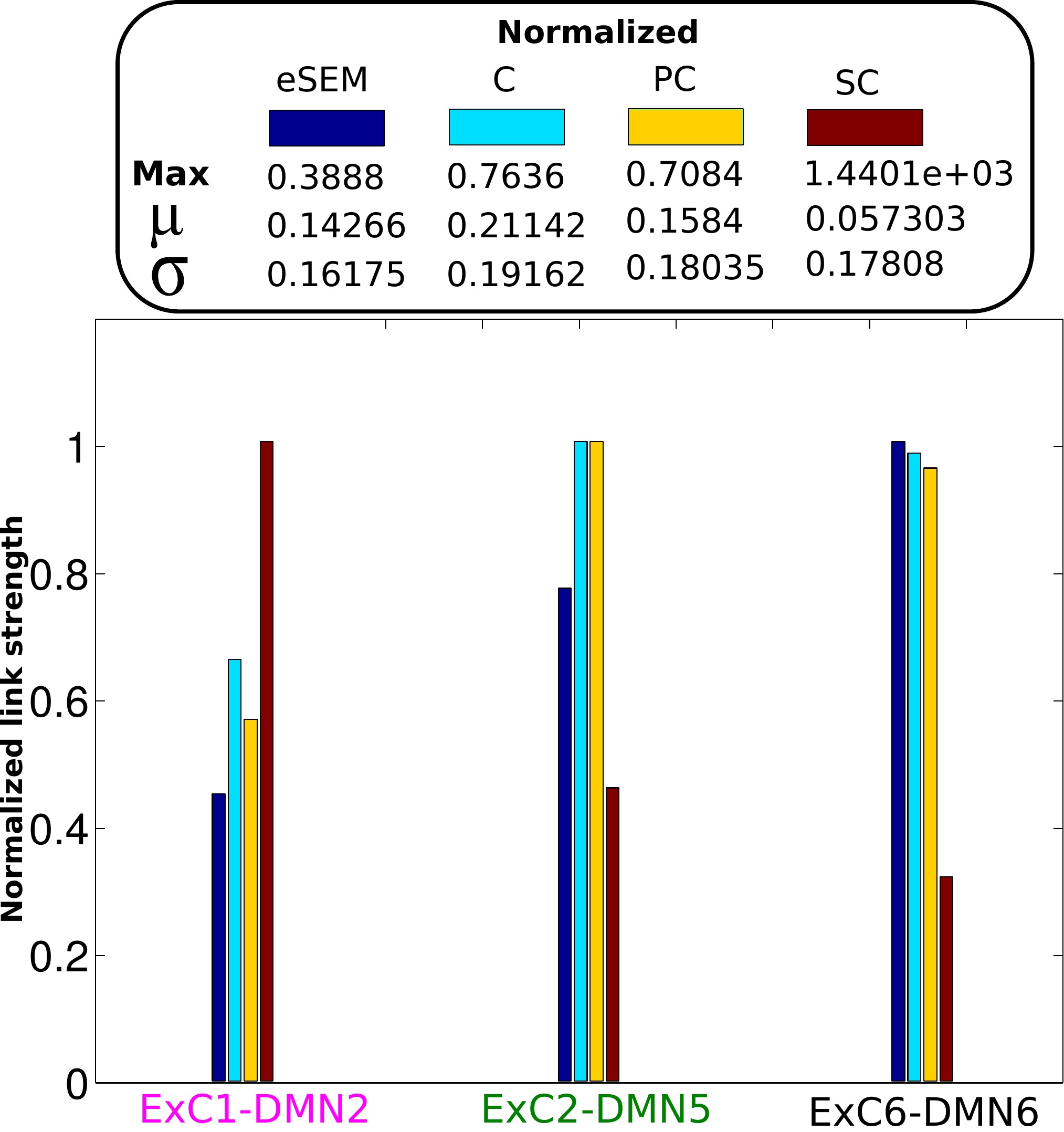}
\caption{\label{fig3} {\bf Connectivity values on specific links}.  All matrices eSEM, C, PC and SC   were normalized in the range $[0,1]$ for visualization purposes. The maximum values used for normalization in each case are shown, as well as the mean ($\mu$) and the standard deviation ($\sigma$)  values for all matrices.
}}
\end{figure}

For a  further analysis we   looked at the values of  SC,  eSEM, C and PC on three specific  links: the ones   with a  highest value in each SC, FC and EC:
\begin{itemize}
\item The    \textit{structural link},    the pair of ROIs sharing the highest value of  SC, which was ExC1-DMN2 (x-label colored in magenta  in figure \ref{fig3}).
\item The    \textit{functional  link},   the pair of ROIs with highest value of C, which was coincident with the pair with maximum PC, that was  ExC2-DMN5  (x-label colored in green).
\item The   \textit{effective link}, the pair  of ROIs with highest value of  SEM:  ExC6-DMN6  (x-label colored in black).  
\end{itemize}
From the structural link, and although the average value of eSEM performed similarly to C (figure \ref{fig2}B),   eSEM gave a significantly smaller value than C and PC, reflecting high  relation  between  ExC1 and DMN2 due to SC.  By looking at the functional link,  eSEM also provided a high value, indicating that   the two areas with neuronal activity most statistical similar  each other, ExC2 and DMN5,  also had  a high directed influence between them. Finally, results on the effective link showed that the link with the highest dynamical influence, from DMN6 to ExC6, also had a high value of C and PC.

\begin{figure}[htb]
\center{\includegraphics[width=18cm]{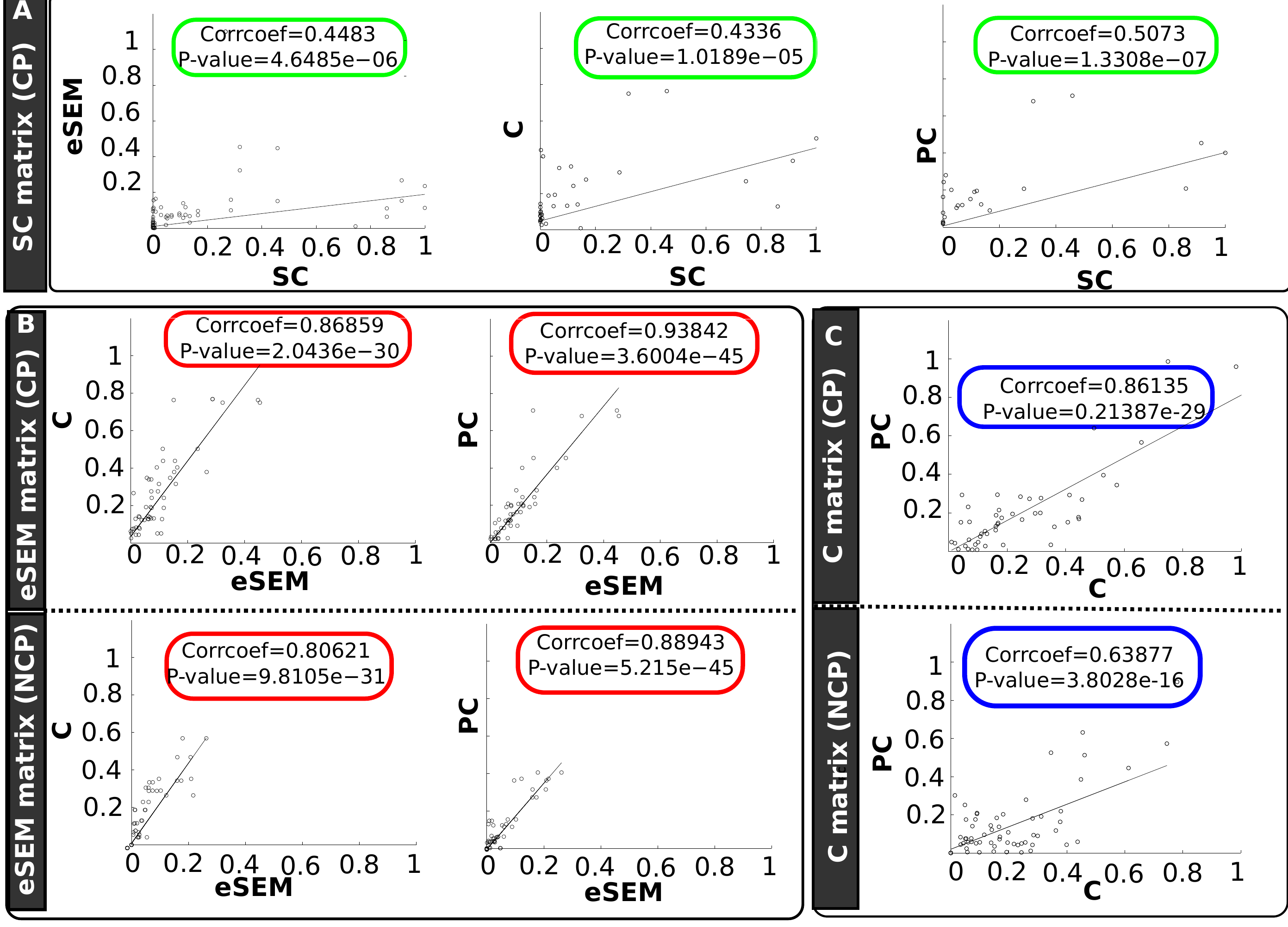}
\caption{\label{fig4} {\bf Scatter plots between different connectivity matrices } and separating in two groups: structurally connected pairs (CP) and  non-connected pairs (NCP).  Different panels are showing scatter plots of \textbf{A:} (green rectangles) SC with eSEM, C and PC,  \textbf{B:} (red rectangles) eSEM with C and PC,  \textbf{C:} (blue rectangles) C with PC.
}}
\end{figure}

\begin{figure}[htb]
\center{\includegraphics[width=18cm]{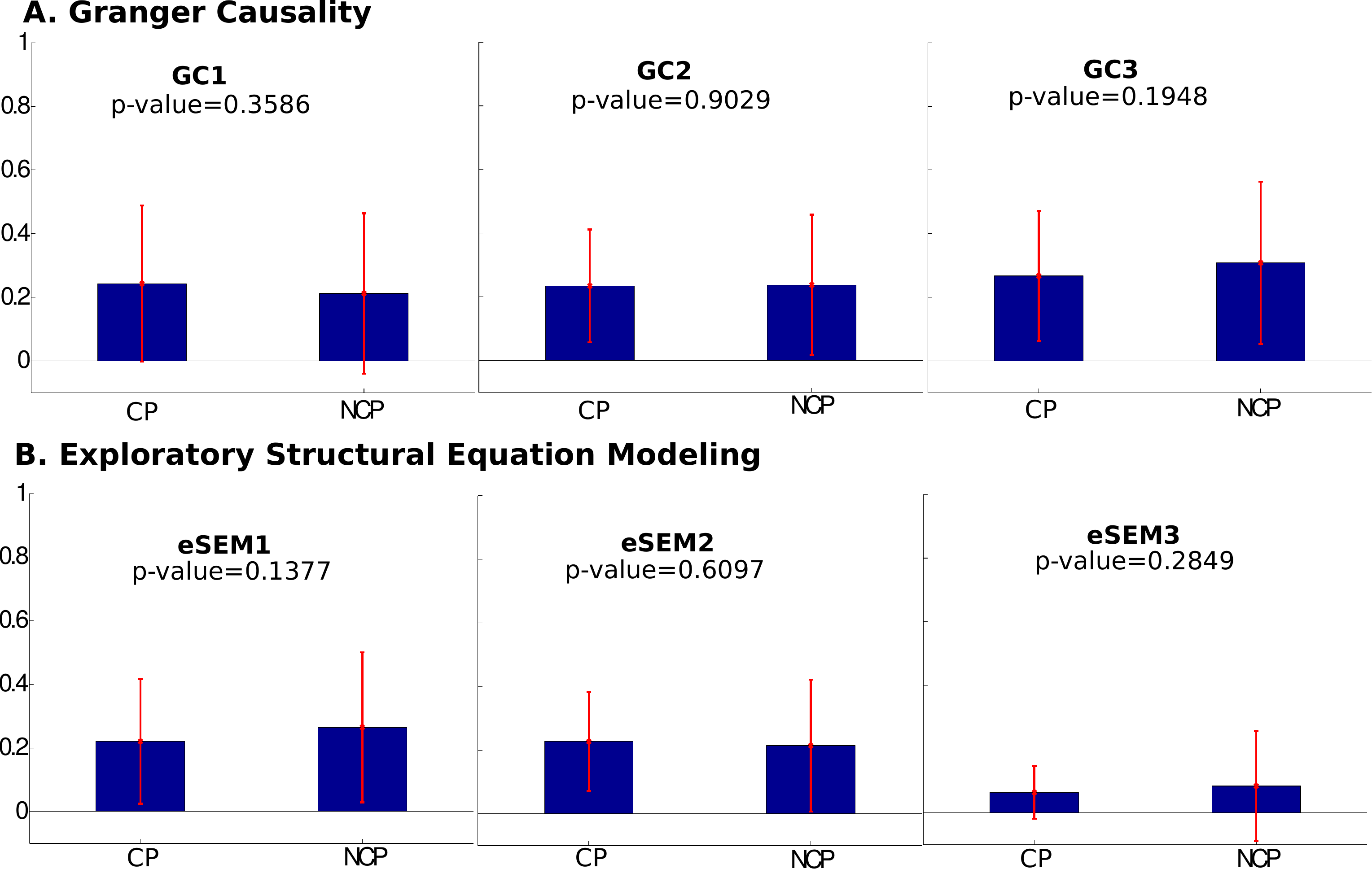}
\caption{\label{fig5} {\bf Mean values of  structurally connected pairs (CP) and not connected pairs (NCP) across several lags in \textbf{A}: Granger Causality and \textbf{B:} eSEM}. eSEM1, eSEM2 and eSEM3 (the same as GC1, GC2 and GC3) refers to lag=\{1,2,3\} for both eSEM and GC. Notice that, in all the cases, the differences found between the two groups were not significant according to the p-value. So, neither eSEM nor GC distinguished between CP and NCP.
}}
\end{figure}

Beyond  results at the level of individual links,  scatter plots  between the different connectivity matrices   (SC, eSEM, C and PC) for all the pairs  are shown in figure \ref{fig4}.  The matrices resulting from  eSEM, C and PC were significantly correlated with the structural one, SC (rounded green rectangles in figure \ref{fig4}).  Correlation coefficients were 0.44, 0.43 and 0.50 for respectively eSEM, C and PC.

We also found that eSEM matrix was highly correlated with C and PC matrices for both CP and NCP  (rounded red circles); indeed, for  CP  the  correlation was equal to   0.86 (for C)  and 0.93 (for PC). Thus, on CP pairs, PC and eSEM  were approximately equivalent to each other.  When looking to NCP, this  correlation between eSEM  and PC went down to 0.88, still a very high value.

Finally, correlation between C and PC matrices were high for both  CP (corr=0.86) and NCP (corr=0.63). This is represented by the  rounded blue rectangles in figure \ref{fig4}.

\begin{table}[!t]
\textbf{\refstepcounter{table}\label{tab1} Table \arabic{table}.}{  ROI size (mm$^3$). }

\processtable{}
{\begin{tabular}{lll}\toprule
 Network &ROIs & size (mm$^3$) \\\midrule
Sensory Motor (SM) Network  
&SM & 194.960\\ 
   
&  &  \\ 
Default Mode Network (DMN)  
&DMN1 & 97.091\\
&DMN2 & 44.115\\
&DMN3 & 28.374\\
&DMN4 & 22.330\\
&DMN5 & 8.343\\
&DMN6 & 10.826\\ 
&  &  \\ 
Executive Control (ExC) 
&ExC1 & 79.956\\
&ExC2 & 43.313\\
&ExC3 & 52.225\\
&ExC4 & 48.483\\
&ExC5 & 15.769\\
&ExC6 & 15.745\\
&ExC7 & 11.723\\
&ExC8 & 32.752\\\botrule
\end{tabular}}{}
\end{table}

\section{Discussion}

Multiple evidence  have shown   brain topology (ie., structure) supporting dynamics (ie., function) and   brain dynamics reinforcing structure via synaptic plasticity (or   punishing it via synaptic prunning), but the precise relationship between the two (structure and function) is still  challenging \citep{damoiseaux2009,park2013}. 

A powerful method to approach this problem at the large-scale brain organization  is to calculate  structural and functional networks and address their  mutual relationships  \citep{park2013}. Following this strategy, here, we calculated  SC, FC and EC for a very specific brain parcellation, with ROIs covering the 	entirety  of three   well-know resting networks, the executive control, the default mode and the sensory motor network.  After this brain division, we obtained 15 different ROIs and by performing to the same subject  two classes of MRI acquisitions  (one structural, one functional) we made  a careful comparison between SC (ie. fiber number connectivity between ROIs), FC (pairwise C and PC connectivities) and EC (by generalizing SEM to its exploratory version  eSEM).

We  have made use of  eSEM for the inference of functional integration; eSEM,   although rooted in  the SEM framework,   is     exploratory and  can      assess influences between brain regions without assuming any implicit model.   We have   studied how much  similar eSEM was to SC, an compared these results  with equal-time correlational analysis by calculating both C and PC, which are the leading methods to estimate FC.

In the first part of this study, our results showed that  eSEM, in addition to C and PC, were  able to significantly separate the set of non-connected pairs in the structural network from the set of connected pairs. Although the PC analysis is slightly the best one in correlating with the strength of structural links, interestingly, for the specific situation of  restricting to connected pairs, the eSEM estimation was  almost identical to PC (correlation value of 0.93). The  fact that  eSEM provided    a similar  correlation with SC   to the one    achieved by   C and PC   makes   the use of eSEM    equally valid as  C and PC for FC brain studies.

 On the other hand it must be stressed that eSEM also provided information about the case of fiber pairs
where information  preferably flowed  in one direction. These results showed the usefulness of fully connected eSEM inference of directed dependencies between structurally connected ROIs in the human brain.

It is important to emphasize that there are other studies  also relating SEM with C and/or PC.  Thus, it was shown that PC performed better than SEM in  identifying  local patterns of interaction detected by SC \citep{Marrelec2009} and that C and PC were suitable candidates to simultaneously analyse SC and FC in the entire brain \citep{Horn2013}; furthermore, this evidence was even stronger when focused on the Default Mode Network, an important RSN with important implications in memory performance. In another study, 
 when SEM was used in combination with DTI  data     \citep{Voineskos2012}, the authors approached  aging and cognitive performance using SC   to analyse tract degeneration and  SEM to address    white matter tract integrity.

 In the second part of our study, we   have applied eSEM and multivariate Granger Causality to show that, when lagged time series are considered to estimate EC, the results are much less correlated    with  SC (figure \ref{fig5}). This suggests that fast interactions (captured by instantaneous measures of connectivity) are shaped by the structural strength, whilst slower directed functional interactions (those captured by methods relying on temporal correlations) are less shaped by the structural
strength. In other words, at slow time scales, the statistical dependencies among ROIs appear to be less related to the details of the underlying structural connectivity.

 The fact that the  lagged methods   found influences between brain regions acting at a  time scale equal to the sampling time  suggests that the lagged algorithms may be seen as complementary to the standard correlational analysis. The  eSEM method, here described, is   suitable tool  to detect   those directed
functional interactions which cannot be described merely to the presence of a strong structural connection between brain areas.

\textit{To summarize,}  based on the evidence  that RSNs are functionally integrated by structural connections  \citep{Heuvel2013} here, by  building  a very simple large-scale brain system consisting of three of those RSNs, and without assuming any implicit connectivity between them, we have shown that eSEM can perform equally well than C and PC in correlating with SC,  thus encouraging the use of eSEM for FC studies at rest. Whether this statement still holds during task paradigms  needs to be investigated.

\section*{Disclosure/Conflict-of-Interest Statement}
The authors declare that the research was conducted in the absence of any commercial or financial relationships that could be construed as a potential conflict of interest.


\section*{Author contributions}

IE and BM performed MRI acquisitions; IE preprocessed and postprocessed the MRI data; CAM performed all simulations;  CAM, ID, LR, YR, DM, SS and JMC developed the methods; CAM, ID, LR, YR, DM, SS and JMC wrote the paper;  DM,  SS and JMC designed the research; all the authors reviewed the manuscript.


\section*{Acknowledgements}

This work has been conducted within the Computational Neuroimaging Lab in the Biocruces Institute, and using the computing resources provided by BCAM.  

\paragraph{Funding\textcolon} 

Financial support from the Basque Government (BERC 2014-2017) and the Spanish Ministry  of Economy and Competitiveness (MINECO) : BCAM Severo Ochoa accreditation (SEV-2013-0323) to  CAM and LR; from Ikerbasque: The Basque Foundation for Science and Euskampus at UPV/EHU to  JMC; from Bizkaia Talent (AYD-000-285) to SS.

 \section*{Data availability}
The SC matrices for each subject and the time series rs-fMRI data are available upon the  reader's request. 

\section*{Supplementary Material}

\subsection*{Effective connectivity (EC)  by multivariate Granger Causality (GC).}  

Let us first describe bivariate Granger causality \citep{granger1969}.
Suppose we model the temporal dynamics of a stationary time series
$\{\xi_n\}_{n=1,.,N+m}$ by an autoregressive model of order $m$:
$$\xi_n=\sum_{j=1}^m A_j\; \xi_{n-j}+E_n,$$
and by a bivariate autoregressive model which takes into account
also a simultaneously recorded time series $\{\eta_n\}_{n=1,.,N+m}$:
$$\xi_n=\sum_{j=1}^m A^\prime_j \;\xi_{n-j}+\sum_{j=1}^m B_j \;\eta_{n-j}+E^\prime_n.$$
The coefficients of the models are calculated by standard least
squares regression;  $m$ is usually selected according to the Akaike
criterion applied to the VAR modeling of the multivariate time series,  providing an    optimal order of the model \citep{Akaike1974}.

It can be said that $\eta$ Granger-causes $\xi$ if the variance of residuals $E^\prime$ is significantly smaller than
the variance of residuals $E$, as it happens when coefficients $B_j$ are jointly significantly different from zero. This can be tested by performing an either  F-test or Levene's test for equality of variances \citep{geweke1982}. An index measuring the strength of the causal interaction is  $\delta=1-{\langle {E^\prime}^2\rangle\over \langle E^2\rangle}$, 
where $\langle \cdot\rangle$ means averaging over $n$ (note that $\langle E\rangle =  \langle E^\prime \rangle =0$). Exchanging the roles of the two time series, one may equally test causality in the opposite direction, i.e. to check whether $\xi$ Granger-causes $\eta$.

In the conditioned case, let $\{\psi^a_n\}_{n=1,.,N+m}$, $a=1,\ldots,M$, be $M$ other simultaneously recorded time series. When several variables are present in the system, the Granger influence $\eta \to \xi$ must take into account their possible conditioning effect.  In this case, it is recommended to treat the data-set as a whole, including the $\psi$ times series in both the autoregressive models for $\xi$ described above.  To assess causality in GC, another VAR is learned from data excluding one variable (the candidate driver) from the input set of variables. Then, an F-test is applied to assess significance of the variance reduction due to the inclusion of the candidate driver variable.    The conditioned Granger causality $\eta \to \xi$  measures the reduction in the variance of residuals going from one to other of the following two conditions: (i) all variables $\psi$ are included in the model and (ii) all variables $\psi$ and the variable $\eta$ are included. Conditioning on the remaining variables allows to discard indirect interactions that would be recognized as direct by the pairwise approach. We refer the reader to \citep{Stramaglia2014} for a discussion on advantages and pitfalls of pairwise and conditioned Granger causality. In this paper we will refer to GC1, GC2, GC3 when discussing the application of GC with m=1,2,3 respectively.


\bibliographystyle{frontiersinSCNS&ENG} 
\bibliography{semBib}



\clearpage
\section*{Figures}

\clearpage

\clearpage

\clearpage

\clearpage

\setcounter{figure}{0}
\renewcommand{\thefigure}{S\arabic{figure}}
\begin{figure}[h]
\centering{
\includegraphics[width=12cm]{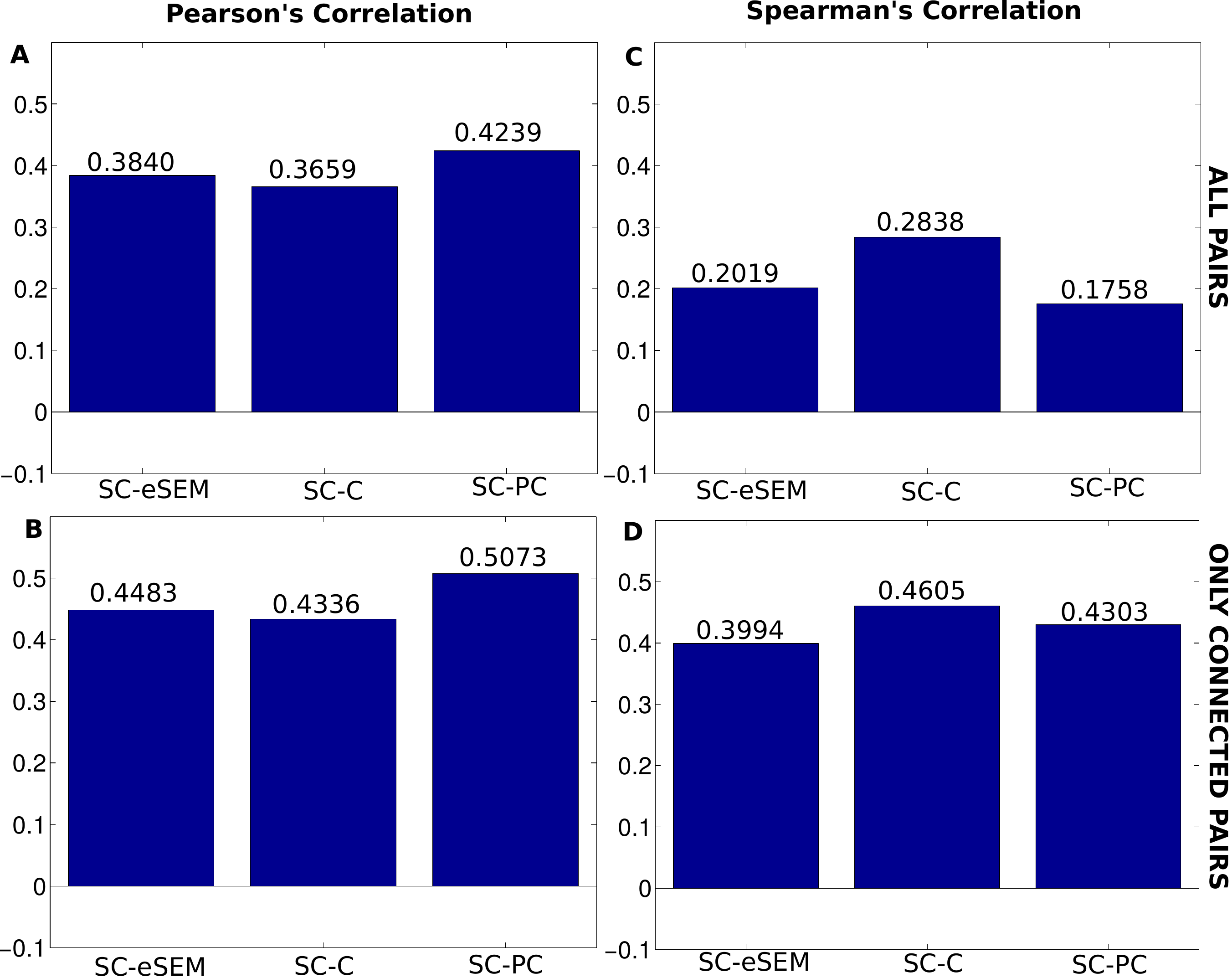}
}
\caption{ {\bf Pearson's vs Spearman's correlations.} The same  figures     \ref{fig2}B$_1$ and \ref{fig2}B$_2$ are now plotted together with  the results from Spearman's correlations for comparison purposes.} 
 \label{figS1}
\end{figure}

 \clearpage
\renewcommand{\thefigure}{S\arabic{figure}}
\begin{figure}[h]
\centering{
\includegraphics[width=12cm]{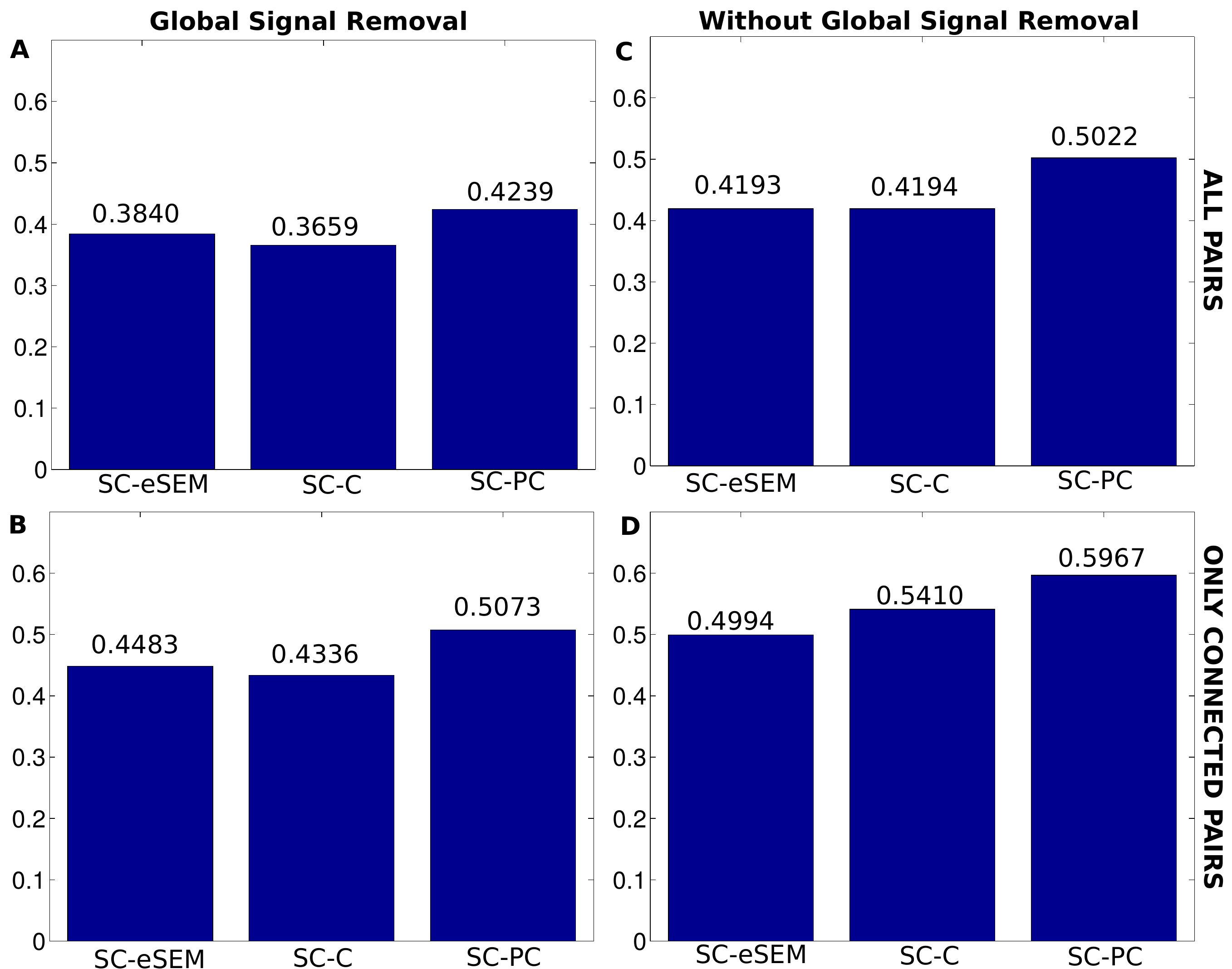}
}
\caption{ {\bf The effect of applying global signal regression vs not appliying it to the time-series rs-fMRI data.} Similar to figures \ref{fig2}B$_1$ and \ref{fig2}B$_2$, here we compared the results of applying global signal removal to the time-series data (panels A and B, and all   other figures in this manuscript), to the results without   global signal removal (panels C and D).}
 \label{figS2}
\end{figure}

\end{document}